\documentclass[12pt,a4paper]{article}

\def\simequiv{\mathop{\simeq}}

\renewcommand{\Re}{{\rm Re}}
\newfont{\blackboard}{msbm10 scaled\magstep2}
\newcommand{\Z}{\mbox{\blackboard\symbol{"5A}}}

\begin{document}
\title{Vacuum Energy Induced by an External Magnetic Field in a Curved Space}
\author{Yu. A. Sitenko\thanks{e-mail: yusitenko@bitp.kiev.ua} 
   \and D. G. Rakityansky\thanks{e-mail: radamir@quantum.bitp.kiev.ua}\\
Bogolyubov Institute for Theoretical Physics\\
National Academy of Sciences of Ukraine}
\date{}
\maketitle

\begin{abstract}
The asymptotic expansion of the product of an operator raised to 
an arbitrary power and an exponential function of this operator is obtained. 
With the aid of this expansion, the density of vacuum energy induced by a 
static externel magnetic field of an Abelian or non-Abelian nature is expressed 
in terms of the DeWitt-Seeley-Gilkey coefficients.
\end{abstract}

\section{Introduction}

It is well known that modern theoretical and mathematical physics widely employs
the asymptotic expansion of the heat kernel,
\begin{equation}\label{1}
<x|e^{-tA}|x>\simequiv_{t\rightarrow 0_+}\sum_{l=0}^\infty E_l(x|A)t^{-\frac{d}{2}+l},
\end{equation}
where $A$ --- is a positive definite elliptic second-order differential operator 
acting in a section of a fiber bundle over a manifold with compact Fiemann base of 
dimension $d$ and where summation is performed over nonnegative integral values 
of $l$. The coefficients of this asymptotic expansion (DeWitt-Seeley-Gilkey 
coefficients \cite{DeWitt,Seel,Gil}) are endomorphisms of a fiber in $x$ and
represent local covariant quantities that are constructed from the coefficient 
functions of the operator $A$, curvature of a fiber and its base, and covariant
derivatives.

In this study, we will consider the diagonal matrix element
\[
<x|A^\alpha e^{-tA}|x>,
\]
where $\alpha$ is a real-valued parameter, and obtain its asymptotic 
expansion for $t\rightarrow 0_+$.  The coefficients in the resulting expansions 
are expressed in terms of the DeWitt-Seeley-Gilkey coefficients. For a 
certain relation between $\alpha$ and $d$, this expansion involves not only 
powers of $t$ but also terms that are logarithmic in $t$.

As an application of the aforementioned result, we consider the problemof 
vacuum energy induced by a static externel magnetic field of an Abelian or 
non-Abelian nature. This problem is studied in spaces of dimensions 
$d\geq 2.$

\section{Asymptotic Expansion of diagonal Matrix element}

We make use of the method based on the symbolic calculus of pseudodifferential
operators \cite{Shub,Wid} and developed in \cite{Gus}.
This method is advatageous in that it can be applied to a wide variety of 
problems and in that it is manifestly covariant under gauge and general coordinate 
transformations. 

We begin by expressing the operator $e^{-tA} (t>0)$ in terms of the resolvent as
\begin{equation}\label{2}
e^{-tA}=\int_C\frac{id\lambda}{2\pi}e^{-t\lambda}(A-\lambda)^{-1},
\end{equation}
where the contour $C$ in the complex $\lambda$ plane circumvents 
counterclockwise the spectrum of the operator $A$, lying on the positive real 
semiaxis. The relation that will serve our present purpose and which is 
analogous to (\ref{2}) has the form
\begin{equation}\label{3}
A^\alpha e^{-tA}=\int_C\frac{id\lambda}{2\pi}\lambda^\alpha
e^{-t\lambda}(A-\lambda)^{-1}.
\end{equation}
If $\alpha\neq n$, where $n\in\Z$ ($\Z$ is the set of integers), the point
$\lambda=0$ is the branch point of the integrand on the right-hand side of 
(\ref{3}). Drawing a cut in the complex $\lambda$ plain along the negative real
semiaxis from the origin to $-\infty$, we can arrange the contour $C$ in such a way
that it lies in the half-plane $\Re\lambda>0$. This can be done because the 
operator $A$ is strictly positive definite; hence, the lower bound of its 
spectrum nonzero: the contour $C$ itresects the real axis at the point lying in
the interval between zero and the lower bound of the spectrum. It follows that 
the matrix element of the operator in (\ref{3}) is represented as an integral
of the resolvent symbol, in just the same way as the matrix element of the 
operator in (\ref{2}) is.

For the purposes of the ensuing analysis, it is more convenient to introduce the 
spectral parameter $\lambda$ in an alternative way. A positive definite operator
$A$ can be represented as
\begin{equation}\label{4}
A=A_0+m^2
\end{equation}
where $m^2>0$ and $A_0$ is a nonnegative operator (the lower bound of its
spectrum is zero, so that zero modes can exist). Apart from (\ref{1}),
there also exists an asymptotic expansion that is associated with the kernel, 
but which corresponds to the operator $A_0$,
\begin{equation}\label{5}
<x|e^{-tA_0}|x>\simequiv_{t\rightarrow 0_+}\sum_{l=0}^{\infty}
E_l(x|A_0)t^{-\frac{d}{2}+l},
\end{equation}
Taking this into account, we can express the operators on  the left-hand sides
of the equations (\ref{2}) and (\ref{3}) in terms of the resolvent of the 
operator $A_0$. This yields
\begin{equation}\label{6}
e^{-tA}=e^{-tm^2}\int_C\frac{id\lambda}{2\pi}e^{-t\lambda}(A_0-\lambda)^{-1},
\end{equation}
\begin{equation}\label{7}
A^\alpha e^{-tA}=e^{-tm^2}\int_C\frac{id\lambda}{2\pi}
(\lambda+m^2)^\alpha e^{-t\lambda}(A_0-\lambda)^{-1},
\end{equation}
where the cut in the complex $\lambda$ plane goes from the point $-m^2$ to 
$-\infty$, and the contour $C$ intersects the real axis at some poiint lying in 
the interval ($-m^2, 0$).
For the diagonal matrix elements of the operators in 
(\ref{6}) and (\ref{7}), we obtain
\begin{equation}\label{8}
<x|e^{-tA}|x>=e^{-tm^2}\int\frac{d^dk}{(2\pi)^d\sqrt{g(x)}}\int_C\frac{id\lambda}
{2\pi}e^{-t\lambda}\sigma^0(x,k;\lambda),
\end{equation}
\begin{equation}\label{9}
<x|A^\alpha e^{-tA}|x>=e^{-tm^2}\int\frac{d^dk}{(2\pi)^d\sqrt{g(x)}}\int_C
\frac{id\lambda}{2\pi}(\lambda+m^2)^\alpha e^{-t\lambda}\sigma^0(x,k;\lambda),
\end{equation}
where $\sigma^0$ is the symbol of the resolvent of the operator $A_0$.
This symbol can be expanded in a series in the powers of homogeneity as
\begin{equation}\label{10}
\sigma^0(x,k;\lambda)=\sum_{l=0}^{\infty}\sigma^0_l(x,k;\lambda),\quad
\sigma^0_l(x,\epsilon k;\epsilon^2\lambda)=
\epsilon^{-2(1+l)}\sigma^0_l(x,k;\lambda),
\end{equation}
where $l$ takes nonnegative integral values. The functions $\sigma^0_l$
satisfy recursion relations. Substituting expansion (\ref{10}) into
(\ref{8}) and performing the required integrations, we arrive at expansion 
(\ref{1}). 

As a result, we obtain 
\begin{equation}\label{23}
<x|A^{\alpha}e^{-tA}|x>\simequiv_{t\rightarrow 0_+}
\sum_{l=0}^\infty E_l(x|A)\frac{\Gamma(\alpha+\frac{d}{2}-l)}{\Gamma(\frac{d}{2}-l)}
t^{-\alpha-\frac{d}{2}+l}+
\end{equation}
\[
+\sum_{l=0}^\infty \sum_{l'=-l\atop l'>\alpha+\frac{d}{2}}^\infty
E_{l+l'}(x|A_0)\frac{\Gamma(l'-\alpha-\frac{d}{2})}{\Gamma(-\alpha-l)\Gamma(l+1)}
m^{2(\alpha+\frac{d}{2}-l')}(-t)^l,\quad\alpha+\frac{d}{2}\neq n;
\]
\begin{equation}\label{24}
<x|A^{n-\frac{d}{2}}e^{-tA}|x>\simequiv_{t\rightarrow 0_+}
\sum_{l=0}^{n-1} E_l(x|A)\frac{\Gamma(n-l)}{\Gamma(\frac{d}{2}-l)}
t^{-n+l}+
\end{equation}
\[
+\sum_{l={\rm max}(n,0)}^\infty E_l(x|A)\frac{\ln(tm^2)^{-1}-\gamma+\psi(l+1-n)
-\psi(\frac{d}{2}-l)}{\Gamma(\frac{d}{2}-l)\Gamma(l+1-n)}(-t)^{l-n}+
\]
\[
+\sum_{l=0}^\infty\quad \sum_{l'={\rm max}(n+1,-l)}^\infty
E_{l+l'}(x|A_0)\frac{\Gamma(l'-n)}{\Gamma(\frac{d}{2}-n-l)\Gamma(l+1)}
m^{2(n-l')}(-t)^l,
\]
where  $n\in\Z$; the notation  ${\rm max}(a,b)$ means that, of the two 
quantities $a$ and $b$, that which takes the larger value must appear in 
the corresponding expression; 
$\psi(z)=\frac{d}{dz}\ln\Gamma(z)$ is the digamma function;
and $\gamma=-\psi(1)$ is the  Euler constant.  The first (finite) sum in the
(\ref{24}) naturally vanishes for $n\leq 0$.

\section{Determination of the renormalized density of induced vacuum energy}

The results of the preceding section can be used to solve the problem of the 
vacuum energy induced by the curvature of a fiber, the strength of a static
external magnetic field. To determine the vacuum-energy density, it is 
necessary to specify the specify regularization and renormalization procedures.
We apply the standard procedure, according to which ultraviolet divergencies
are regularized by introducing a cuttoff factor.

Let us consider a static $(d+1)$-dimensional space-time endowed with a metric 
specified by the relation
\begin{equation}\label{27}
ds^2=-(dx^0)^2+g_{\mu\nu}(x)dx^\mu dx^\nu, \quad \mu,\nu=1,...,d,
\end{equation}
where $x^0$ is the time. We write Dirac equation in the form 
\begin{equation}\label{28}
(-i\partial_0+H_D)\psi=0,
\end{equation}
where
\begin{equation}\label{29}
H_D=-i\alpha^\mu(x)\nabla_\mu + \beta m
\end{equation}
is the Dirac Hamiltonian.
For a spinor field, the zeroth component of energy-momentum tensor has the form
\begin{equation}\label{32}
T_{00}=\frac{i}{2}[\psi^+(\partial_0\psi)-(\partial_0\psi^+)\psi].
\end{equation}
For the vacuum energy of the second-quantized field, we obtain the expression.
\begin{equation}\label{34}
\int d^dx\sqrt{g(x)}\varepsilon(x)=-\frac{1}{2}\sum_\omega|\omega|,
\end{equation}
where $g={\rm det}g_{\mu\nu}$, and $\varepsilon$ is the vacuum-energy density.
Expression  (\ref{34}) is not well-defined, because the sum on the right-hand
side can diverge for $\omega\rightarrow\pm\infty$ (so-called ultraviolet
divergence).  Supplementing the summand with a factor that decreases 
exponentially for $\omega\rightarrow\pm\infty$, we define a regularized 
vacuum energy as
\begin{equation}\label{35}
\int d^dx\sqrt{g(x)}\varepsilon^{\rm reg}(x)=-\frac{1}{2}\sum_\omega
|\omega|e^{-t\omega^2}.
\end{equation}

Let us represent the Klein-Gordon equation in the form
\begin{equation}\label{36}
(\partial_0^2+H_S)\phi=0
\end{equation}
where
\begin{equation}\label{37}
H_S=-\Delta+\xi R(x)+m^2,
\end{equation}
For a scalar field, the zeroth component of the canonical energy-momentum 
tensor is given by
\begin{equation}\label{38}
T_{00}=(\partial_0\phi^*)(\partial_0\phi)+(\nabla_\mu\phi^*)(\nabla^\mu\phi)
+(\xi R+m^2)\phi^*\phi;
\end{equation}
For the vacuum energy of a second-quantized scalar field, we then obtaine the 
formal expression
\begin{equation}\label{40}
\int d^dx\sqrt{g(x)}\varepsilon(x)=\sum_{\omega>0}\omega
\end{equation}
Accordingly, the regularized expression has the form
\begin{equation}\label{41}
\int d^dx\sqrt{g(x)}\varepsilon^{\rm reg}(x)=\sum_{\omega>0}\omega
e^{-t\omega^2}.
\end{equation}

Thus, the regularized vacuum-energy density can be represented as
\begin{equation}\label{42}
\varepsilon^{\rm reg}(x)=-\frac{1}{2}{\rm tr}<x||H_D|\exp(-tH_D^2)|x>
\end{equation}
for a spinor field and as
\begin{equation}\label{43}
\varepsilon^{\rm reg}(x)={\rm tr}<x|\sqrt{H_S}\exp(-tH_S)|x>
\end{equation}
for a scalar field

By using expansions (\ref{23}) and (\ref{24}) at $\alpha=\frac{1}{2}$,
we obtain
\begin{equation}\label{44}
<x|A^{1/2}e^{-tA}|x>\simequiv_{t\rightarrow 0_+}\frac{\sqrt{\pi}}{2}E_0(x|A_0)
t^{-3/2}-
\end{equation}
\[
-\frac{1}{\pi}\sum_{l=0}^\infty\sum_{l'=2}^\infty
E_{l+l'}(x|A_0)\frac{\Gamma(l'-3/2)\Gamma(l+3/2)}{\Gamma(l+1)}m^{3-2l'}t^l,
\quad d=2,
\]
\begin{equation}\label{45}
<x|A^{1/2}e^{-tA}|x>\simequiv_{t\rightarrow 0_+}\frac{2}{\sqrt{\pi}}E_0(x|A_0)
t^{-2}-
\end{equation}
\[
-\frac{1}{\sqrt{\pi}}[m^2E_0(x|A_0)-E_1(x|A_0)]t^{-1}-
\]
\[
-\frac{1}{\pi}\sum_{l=2}^\infty\sum_{l'=0}^\infty
E_{l'}(x|A_0)\frac{\Gamma(l-1/2)}{\Gamma(l-l'+1)\Gamma(l-1)}
[\ln(tm^2)^{-1}-\gamma+\psi(l-1)-
\]
\[
-\psi(l-1/2)](-m^2)^{l-l'}t^{l-2}-
\frac{1}{\pi}\sum_{l=0}^\infty\sum_{l'=3}^\infty
E_{l+l'}(x|A_0)\frac{\Gamma(l'-2)\Gamma(l+3/2)}{\Gamma(l+1)}
m^{4-2l'}t^l,\quad d=3,
\]
\begin{equation}\label{46}
<x|A^{1/2}e^{-tA}|x>\simequiv_{t\rightarrow 0_+}\frac{3\sqrt{\pi}}{4}E_0(x|A_0)
t^{-5/2}-
\end{equation}
\[
-\frac{\sqrt{\pi}}{2}[m^2E_0(x|A_0)-E_1(x|A_0)]t^{-3/2}-
\]
\[
-\frac{1}{\pi}\sum_{l=0}^\infty\sum_{l'=3}^\infty
E_{l+l'}(x|A_0)\frac{\Gamma(l'-5/2)\Gamma(l+3/2)}{\Gamma(l+1)}
m^{5-2l'}t^l,\quad d=4.
\]
Introducing the notation
\begin{equation}\label{47}
H_{D,0}=H_D|_{m=0},\quad H_{S,0}=H_S|_{m=0},
\end{equation}
we can see that, both in the case of
\begin{equation}\label{48}
A_0=H^2_{D,0},
\end{equation}
and in the case of
\begin{equation}
A_0=H_{S,0}
\end{equation}
the operator $A_0$ has the form
\begin{equation}\label{50}
A_0=-\Delta+X(x),
\end{equation}
where
\begin{equation}\label{51}
X(x)=\frac{i}{4}[\alpha^\mu(x),\alpha^\nu(x)]_-F_{\mu\nu}(x)
+\frac{1}{4}R(x)
\end{equation}
in the former case and
\begin{equation}\label{52}
X(x)=\xi R(x)
\end{equation}
in the latter case; in (\ref{51}), we introduce the notation
\begin{equation}
F_{\mu\nu}(x)=\partial_\mu V_\nu(x)-\partial_\nu V_\mu(x)-i[V_\mu(x), V_\nu(x)]_-.
\end{equation}

\section{Resume}

We have obtained the asymptotic expansions (\ref{23}) and (\ref{24})
for the diagonal matrix element of the operator product $A^{\alpha}e^{-tA}$
for $t\rightarrow 0_+$. At $\alpha=-\frac{d}{2}+n$, where $n\in \Z$, the 
relevant expansion envolves not only powers of $t$ but also terms that are 
logarithmic in $t$.

We have also solved the problem of vacuum energy induced by a static magnetic 
external magnetic field in a curved space.

\end{document}